\documentclass[floatfix,twocolumn,showpacs,preprintnumbers,amsmath,amssymb,pra,superscriptaddress,longbibliography]{revtex4-1}

\usepackage{color}
\usepackage[usenames,dvipsnames,svgnames,table]{xcolor}
\usepackage[colorlinks=true,linkcolor=blue,urlcolor=blue,citecolor=blue]{hyperref}
\usepackage{mathtools}
\usepackage{graphicx}
\usepackage{dcolumn}
\usepackage{array}
\usepackage{lipsum}
\usepackage{bm}
\usepackage{subfigure}
\usepackage{amssymb}
\usepackage{multirow}
\usepackage{tabularx}
\usepackage{amsmath}
\usepackage{braket}
\usepackage{csquotes}
\usepackage{lipsum}
\usepackage{mathrsfs}
\usepackage{MnSymbol}
\usepackage{overpic}

\newcommand{\beq}{\begin{equation}}
\newcommand{\eeq}{\end{equation}}
\newcommand{\bea}{\begin{eqnarray}}
\newcommand{\eea}{\end{eqnarray}}
\newcommand{\beal}{\begin{align}}
\newcommand{\eeal}{\end{align}}

\begin{document}

\title{Van-der-Waals exchange-correlation functionals and their high pressure and warm dense matter applications}

\author{Jan Vorberger}
\email{j.vorberger@hzdr.de}
\affiliation{Helmholtz-Zentrum Dresden-Rossendorf e.V. (HZDR), 01328 Dresden, Germany}

\author{Gabriel J. Smith}
\affiliation{Department of Chemistry, Michigan State University, 48824 East Lansing (MI), USA}

\author{William Z. Van Benschoten}
\affiliation{Department of Chemistry, Michigan State University, 48824 East Lansing (MI), USA}

\author{Hayley R. Petras}
\affiliation{Department of Chemisty, University of Iowa, 52242, Iowa City (IA), USA}

\author{Zhandos Moldabekov}
\affiliation{Helmholtz-Zentrum Dresden-Rossendorf e.V. (HZDR), 01328 Dresden, Germany}

\author{Tobias Dornheim}
\affiliation{Helmholtz-Zentrum Dresden-Rossendorf e.V. (HZDR), 01328 Dresden, Germany}
\affiliation{Center for Advanced Systems Understanding (CASUS), D-02826 G\"orlitz, German}

\author{James J. Shepherd}
\affiliation{Department of Chemistry, Michigan State University, 48824 East Lansing (MI), USA}



\begin{abstract}
We investigate basic hydrogen quantities like the molecular bond length, the molecular dissociation energy and the van-der-Waals interaction in idealized situations in an effort to discern a suitable exchange-correlation functional for the molecular to metal transition in warm dense hydrogen. The best reproduction of bond length and dissociation energy is given by the r$^2$SCAN functional, several vdW functionals and also HSE06 fair qualitatively and quantitatively no better than PBE or worse. r$^2$SCAN however fails at reproducing the van-der-Waals energy minimum. In addition we investigate quantities like the static and dynamic ion structure factor, and the electronic DOS to determine differences between exchange-correlation functionals with and without van-der-Waals corrections in the transition region from the molecular to the metallic regime of hydrogen.
\end{abstract}

\maketitle
\section{Introduction}
Simulations of dense hydrogen and its properties have come a long way in recent years~\cite{wdm_book,vorberger2025roadmapwarmdensematter,bonitz_pop_2024,Bohme_PRL_2022,Dornheim_MRE_2024}.
Phase transitions and the nature of ionization and dissociation in hydrogen due to high pressure or warm dense conditions, respectively, are of high interest because they influence planetary isentropes or fusion compression paths~\cite{Helled2020b,preising_ApJS_2023,icf-collab_prl_24,vorberger2025roadmapwarmdensematter}. For example, the location of the molecular to metal transition in dense hydrogen influences the thermal expansion coefficient and  the Gr\"uneisen parameter in the interior of Saturn leading to stable layers, even when the isentropes do not cross the  liquid-liquid phase transition (LLPT) line~\cite{preising_ApJS_2023}. Predictions for the location of the molecular to metal transition in the high pressure fluid phase have changed over the years and the most recent models rely on  density functional theory (DFT) simulations.
The latest work on the LLPT in hydrogen shows attempts to improve on the LLPT phase transition line using more advanced exchange-correlation functionals (xc-functionals) in the density functional molecular dynamics (DFT-MD) simulations~\cite{Knudson_Science_2015,Pierleoni_PNAS_2016,Lu_2019,Bergermann_2024}.

Even though the development of new and advanced xc functionals should be from first principles~\cite{Perdew_prescr_2005}, by necessity and convenience it relies also on fitting to experimental and theoretical benchmarks  \cite{Narbe_2018}. One class of such advanced xc-functionals includes non-local van-der-Waals (vdW) corrections that are intended to provide improvements by taking into account rest-charge \& polarization interactions of otherwise neutral objects like atoms or molecules~\cite{GRAY20241,Dion_2004,Lee_2010,Klimes_2010}. We focus here on  non-local vdW functionals because they have been used in the description of warm dense hydrogen~\cite{Knudson_Science_2015}. In addition, we mention atom-pairwise vdW methods ~\cite{Grimme_2006,Grimme_2010,Grimme_2011,Tkachenko_2009} and many-body dispersion methods~\cite{tkachenko_2012} that we however not test here.

At high pressure conditions but low temperatures, i.e., in the solid phases of hydrogen, vdW functionals have shown some improvements compared to. e.g., PBE~\cite{Clay_2014,Morales_qe_2013,Morales_towards_2013}. Undoubtedly, vdW functionals have delivered LLPT curves very close to experimental measurements~\cite{Knudson_Science_2015}.
However, simple estimates of the energy scale of a vdW interaction (in the few meV range), molecular dissociation energies (around $4.5$~eV for the hydrogen molecule) or even the ionisation energy ($13.6$~eV) might question why such a tiny force like the vdW interaction can alter the hydrogen equation of state (EOS) by $100$~GPa in the LLPT range ($1100$~K corresponds to $95$~meV, way more than the vdW energy of $\sim 10$~meV or less) and whether the vdW contributions are the cause for the change. In general, there is the question whether vdW functionals offer any advantages for the description of high energy density applications and warm dense matter. 

Instead of using just another xc-functional to compute another EOS table or LLPT curve, respectively, we here take a simpler approach and check first how basic quantities like the bond length or the dissociation energy, or indeed the vdW interaction are described using several different xc functionals. This follows the spirit of many DFT research programs during which one first checks how a setup of pseudopotentials and xc functionals can reproduce, e.g., lattice constants, surface adhesion energies, or band gaps before then proceeding to calculate more involved or derived quantities like the EOS, conductivities or electron-phonon coupling~\cite{Scholl_2022}. 

In addition to three vdW type functionals (labelled VDW-DF1~\cite{Dion_2004}, VDW-DF2~\cite{Lee_2010}, VDW3-DF3 (optPBE-VDW)~\cite{Klimes_2010}, we perform the same tests with a representative of a Meta-GGA functional, namely r$^2$SCAN~\cite{Furness_2020}, and a hybrid functional, i.e., HSE06~\cite{Krukau_2006}. We benchmark DFT results to experimentally determined bond lengths and dissociation energies and compare vdW interaction curves obtained from DFT with highly accurate quantum Monte Carlo (QMC) results~\cite{cleland_communications_2010}. We also compute pair correlation functions, static structure factors, effective bond lengths and dynamic structure factors in order to follow behavior and properties of different xc functionals from the low density hydrogen gas to hydrogen at high pressure and warm dense matter conditions.

\section{Methods}
We perform DFT and DFT-MD simulations using the \texttt{Vienna Ab-initio Simulation Package} (VASP)~\cite{Kresse_1993,Kresse_1994,kresse1996efficient,kresse1996efficiency}. We generally employ the Mermin formulation of thermal DFT~\cite{Mermin_1965} and the bare Coulomb (pseudo-)potentials as provided with VASP~\cite{blochl1994projector,kresse1999ultrasoft}. Exceptions using the hard PAW pseudopotential are mentioned explicitly. This requires a plane wave cutoff of $6000$~eV. The xc contributions are taken either in LDA~\cite{PW_1992}, PBE~\cite{Perdew_PRL_1996}, VDW-DF1~\cite{Dion_2004}, VDW-DF2~\cite{Lee_2010}, VDW-DF3~\cite{Klimes_2010}, HSE06~\cite{Krukau_2006}, or r$^2$SCAN~\cite{Furness_2020}. Any $k$-point sampling was done using grids of Monkhorst-Pack style~\cite{Monkhorst_1976} and DFT-MD simulations were done in the $NVT$ ensemble using a Nose-Hoover thermostat~\cite{Nose_1984,Hoover_1985}. The DFT-MD simulations feature a supercell with $N=64$ protons. Finite size effects do not play a role in our investigations since we want to compare the effects of different xc functionals. For the calculations of the interaction of just two isolated hydrogen molecules, the supercell has an edge length of $L=20\,$\AA~in order to avoid spurious interactions with periodic images of the molecules. 
For the calculations of the ion dynamic structure factors, we also conduct micro-canonical simulations starting from fully equilibrated $NVT$ simulation snapshots in order to check for effects of the thermostat on the ion or molecule dynamics, respectively. As we are considering an isotropic fluid, the given data for a certain wavenumber are averages over all wavevectors with the same given magnitude.

We calculated near-exact benchmarks for hydrogen molecule dimers. 
These were calculated using the initiator full configuration interaction quantum Monte Carlo (\emph{i}-FCIQMC)~\cite{cleland_communications_2010} method, performed using the \texttt{HANDE-QMC}~\cite{spencer_handeqmc_2019} open-source software.
The \emph{i}-FCIQMC correlation estimates are found using a reblocking procedure applied to data collected within the variable-shift phase for the simulation.~\cite{booth_fermion_2009}
The variable-shift phase is activated once a population of walkers $N_\mathrm{w}\in\{10^5,10^6\}$ is achieved on the reference determinant.
Integral files required by \texttt{HANDE-QMC} are generated using a restricted Hartree--Fock (RHF) calculation performed within the \texttt{Molpro}~\cite{werner_molpro_2012, werner_molpro_2020, werner_molpro_package} software which prepares the necessary \texttt{FCIDUMP}~\cite{knowles_determinant_1989} file used by \texttt{HANDE-QMC}.
When \emph{i}-FCIQMC is fully converged with walker number, the exact energy within the basis is estimated within the stochastic error bar. 
These calculations were performed on isolated molecules taking advantage of the correspondence between isolated molecular calculations and a supercell calculation that has been made sufficiently large.
In order to check that molecules were appropriate, we performed molecular DFT calculations in Q-Chem 5.3~\cite{qchem_software_2021}. 
Observations of the energy-distance curves matched those that are present in this manuscript, supporting our prior claim that the impact from periodic image interactions are very minimal, and justifying our use of a molecular model for the QMC calculations. The QMC results presented in Figs. \ref{fig:h2h2:1}–\ref{fig:h2h2:4} had an average error of $0.00009\%$, corresponding to standard errors no greater than 0.12 meV.

We perform coupled cluster singles, doubles, and perturbative triples (CCSD(T))~\cite{hampel_comparison_1992, deegan_perturbative_1994} calculations using \texttt{Molpro} to generate a complete basis set (CBS) correction for our QMC results~\cite{cleland_study_2011}. The CCSD(T) calculations use correlation consistent basis sets, ranging from double to sextuple polarization (cc-pV$X$Z; $X=$ D, T, Q, $5$, $6$)~\cite{dunning_ccpvdz, dunning_ccpv6z}. We extrapolate the CCSD(T) energies to the CBS limit as a function of $1/X^3$  for the two largest basis sets~\cite{kutzelnigg_rates_1992}, and the CBS correction is taken as the differences between the cc-pVTZ and extrapolated CCSD(T) energies. We then add the CBS energy correction to the cc-pVTZ QMC results to correct them to the CBS limit. For example, the CBS correction accounts for $-0.12$~eV of the $-63.93$~eV total energy of the geometry 1 minimum.
\section{Results}
The following subsections discuss the results of different xc functionals for the hydrogen molecule bond length and dissociation energy (Sec.~\ref{bl}), for the idealized situation of the interaction of two hydrogen molecules (Sec.~\ref{im}), for the EOS (Sec.~\ref{vdweos}), for the static structure in hydrogen across the LLPT (Sec.~\ref{vdwsk}), for the dynamic structure (Sec.~\ref{vdwskw}), and for the electronic DOS (Sec.~\ref{vdwdos}).
\subsection{Hydrogen molecule bond length and dissociation energy}\label{bl}
\begin{figure}[t]
    \centering
    \includegraphics[width=0.5\textwidth,clip=true]{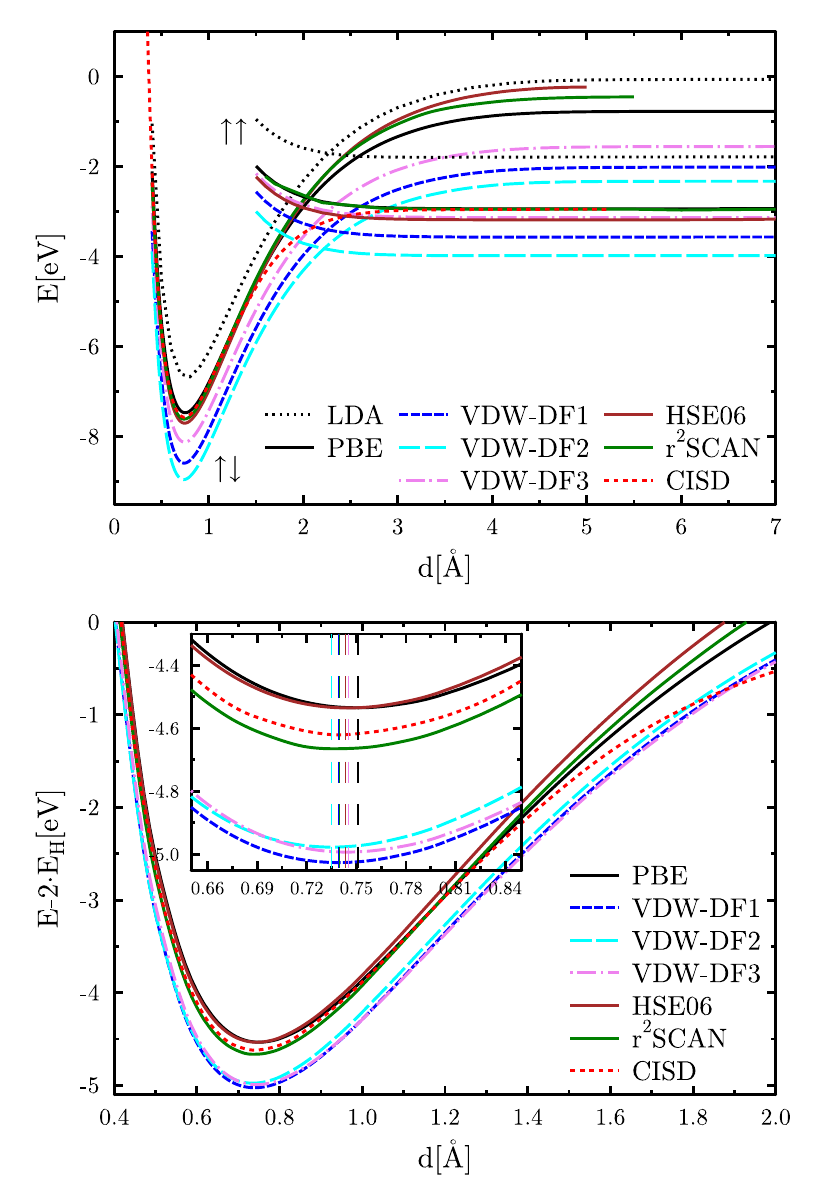}
    \caption{(Top) H$_2$ dissociation curves for different xc functionals. QMC result (CISD) taken from Ref.~\cite{Bonfim2017}. (Bottom) Close up of the potential curve minimum for the H$_2$ molecule. The bond lengths as predicted by DFT are shown as dashed vertical lines in the inset.}
    \label{fig:h2:m2a}
\end{figure}

\begin{table*}[t]
    \centering
    \caption{H$_2$ bond length and dissociation energies}
    \begin{tabular}{c|c|c|c|c|c|c|c|c}
    & LDA & PBE & VDW-DF1 & VDW-DF2 & VDW-DF3 &HSE06 & R2SCAN & exact \\\hline\hline
    bond length $[\AA]$     &  0.766 & 0.750 & 0.739 & 0.735 & 0.745 & 0.7433 & 0.740 & 0.741~\cite{NIST} \\\hline
    dissoc. energy [eV]     & 4.903 & 4.532 & 5.027 & 4.977 & 4.993 & 4.527 & 4.675 & 4.747~\cite{Hoggan_2010} \\\hline
    \end{tabular}
    \label{tab:h2bond}
\end{table*}
As we can see in Fig.~\ref{fig:h2:m2a} and Table~\ref{tab:h2bond}, LDA and PBE overestimate the bond length in a hydrogen molecule by $0.25\,$\AA~and $0.1\,$\AA, respectively. The vdW functionals are closer than PBE to the experimental bond length value~\cite{NIST}, with VDW-DF1 and VDW-DF2 underestimating it and VDW-DF3 overestimating it by a tiny amount. Best results for the bond length can be obtained when using the HSE06 or even better the r$^2$SCAN functional with only a deviation of $0.001\,$\AA. 

PBE underestimates the dissociation energy by about $0.21$~eV compared to experiment~\cite{Kolos_1960,Hoggan_2010}. The three different tested vdW functionals overestimate the dissociation energy by $\sim 0.25$~eV. Best again is the r$^2$SCAN functional with only a deviation of $\sim 0.07$~eV. The result for the dissociation energy using the HSE06 functional is remarkably close to the PBE value.

In the case of pressure dissociation, the different dissociation energies directly influence the LLPT line by virtue of the volume work they require. From
\beq 
W=\int\limits_{V_1}^{V_2}p\ \textnormal{d}V
\eeq
we can calculate, given the hydrogen EOS, what pressure change would be needed to overcome an additional $\sim 0.3 \ldots 0.4$~eV dissociation energy per molecule (comparing, e.g., PBE to any vdW functional). At $T=1500$~K, the required pressure differential is $70\ldots 100$~GPa, which is in agreement with the shift one observes in the simulations between the PBE LLPT line and vdW LLPT lines. 

The shape of the molecular binding and dissociation curve, resp., around the minimum is basically identical between the different xc functionals. A (harmonic) fit of these potential minima gives very much the same fit parameter. There is no difference between the CISD and DFT results in this regard.

From these considerations, there is no clear evidence that vdW-type xc functionals should give superior results to PBE. They represent a situation with similar deviation from experiment but opposite sign than PBE. r$^2$SCAN and HSE06 perform slightly better and r$^2$SCAN should be preferred for the calculation of the EOS and LLPT if one chooses to solely base this decision on the results for bond length and dissociation energy.
\subsection{Interaction of two isolated hydrogen molecules}\label{im}
\begin{figure*}[t]
    \centering
    \begin{overpic}[width=\textwidth]
     {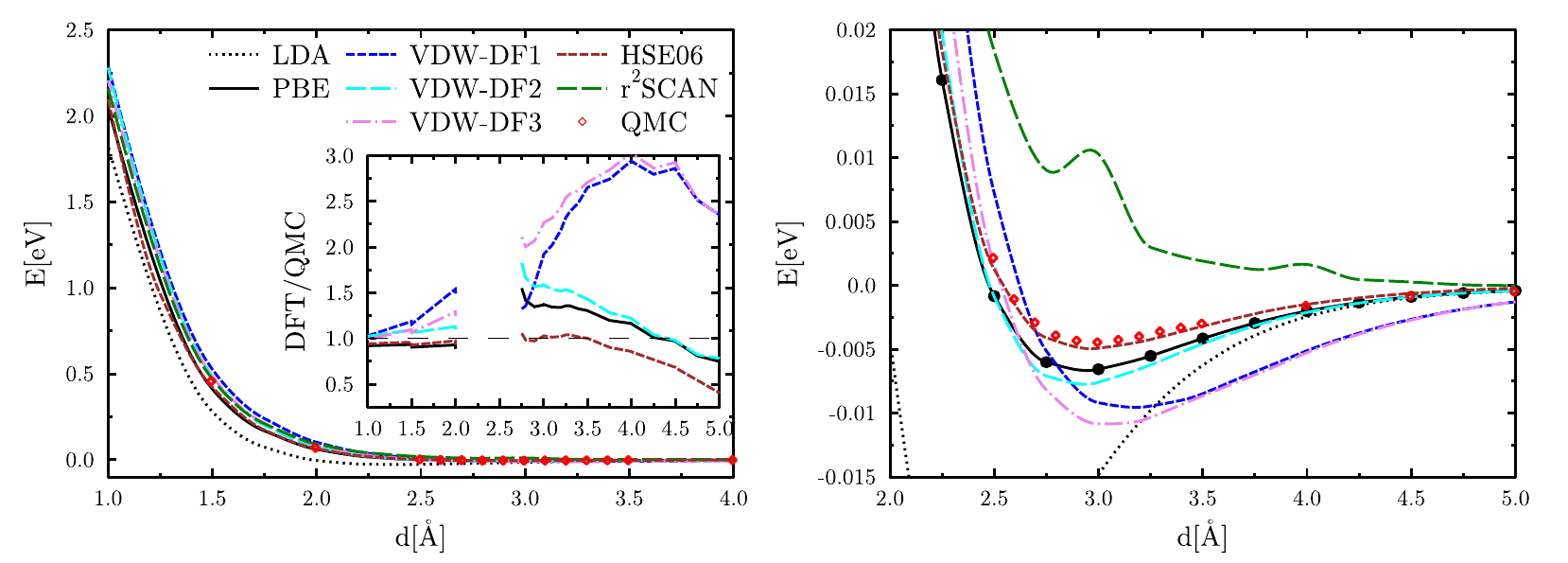}
     \put(85,19){\includegraphics[scale=0.5]{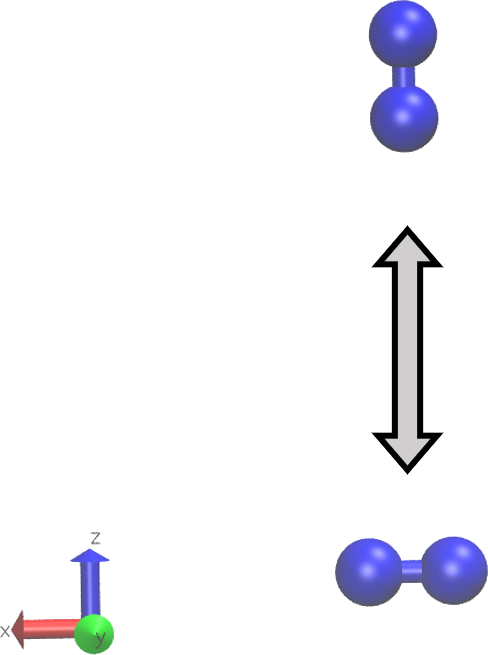}}
     \put(93.2,19){a}
     \put(92,25){d}
    \end{overpic}
    \caption{Model interaction geometry 1. The relative position of the four atoms ($L\ldots$ edge length of supercell, $d\ldots$ distance between molecules, $a\ldots$ molecule bond length) is shown in the top right corner: $(L/2,L/2,L/2-d/2)\,,(L/2+a,L/2,L/2-d/2)\,,(L/2+a/2,L/2,L/2+d/2)\,,(L/2+a/2,L/2,L/2+d/2+a)$ The left panel contains the overview and as the inset the relative deviation of the DFT results (featuring different xc functionals) from the QMC result (no curves are given in the inset for distances where the QMC energy is zero). The right panel shows the area of the vdW minimum of the interaction energy of the system of two molecules as function of the distance of their centre of mass. The energy at maximum intermolecular distance was taken as the zero point of the energy. QMC results contain standard errors no greater than 0.05 meV (smaller than the symbol size).}
    \label{fig:h2h2:1}
\end{figure*}
Next, we would like to study the polarisation and vdW effects between different hydrogen molecules in isolation. In order to do so, we place two hydrogen molecules featuring a bond length optimized using the respective xc functional in an otherwise empty box (supercell). We vary the relative orientations of these two molecules and call the possible arrangements geometries. Geometry 1 is given in Fig.~\ref{fig:h2h2:1}. The distance between the centres of mass is varied and the resulting energies are plotted in the same figure as well.

We observe a repulsion between the two molecules for distances $d$ smaller than $2.5\,$\AA~and a vanishing energy value for large intermolecular spacing (the value of the energy at maximum separation was subtracted from all energies). For separations in the vicinity of $3\,$\AA, a minimum in the energy is observed. This attraction is due to charge fluctuations and is usually called the vdW effect. It is the reason for the introduction of specific vdW xc functionals that were intended to improve forces and energies in such situations where atoms and molecules interact. 

We take the red QMC curve (our own i-FCIQMC result) as the exact benchmark. The LDA curve shows such strong deviations in magnitude and location of this energy minimum when compared to other DFT and the QMC results that we will discard all the LDA results in the discussion. Of the more reasonable results, PBE is quite close to the QMC curve. The minima of the energy of QMC and PBE are at the very similar distances near $d=2.95\,$\AA, however, the QMC minimum is shallower by about $25\%$ than the PBE one. Of the three different vdW functionals, only VDW-DF2 is close to PBE but worse than PBE when compared to QMC. Both VDW-DF1 and VDW-DF3 show a minimum that is too deep and a location of said minimum at too large a distance.

For this geometry, the HSE06 curve is almost indistinguishable from the QMC curve. In stark contrast, the r$^2$SCAN curve does not show a minimum like the other xc functionals and thus lacks a proper description of vdW effects. From now on, we thus drop the r$^2$SCAN functional from our investigation.

\begin{figure*}[t]
    \centering
    \begin{overpic}[width=\textwidth]
     {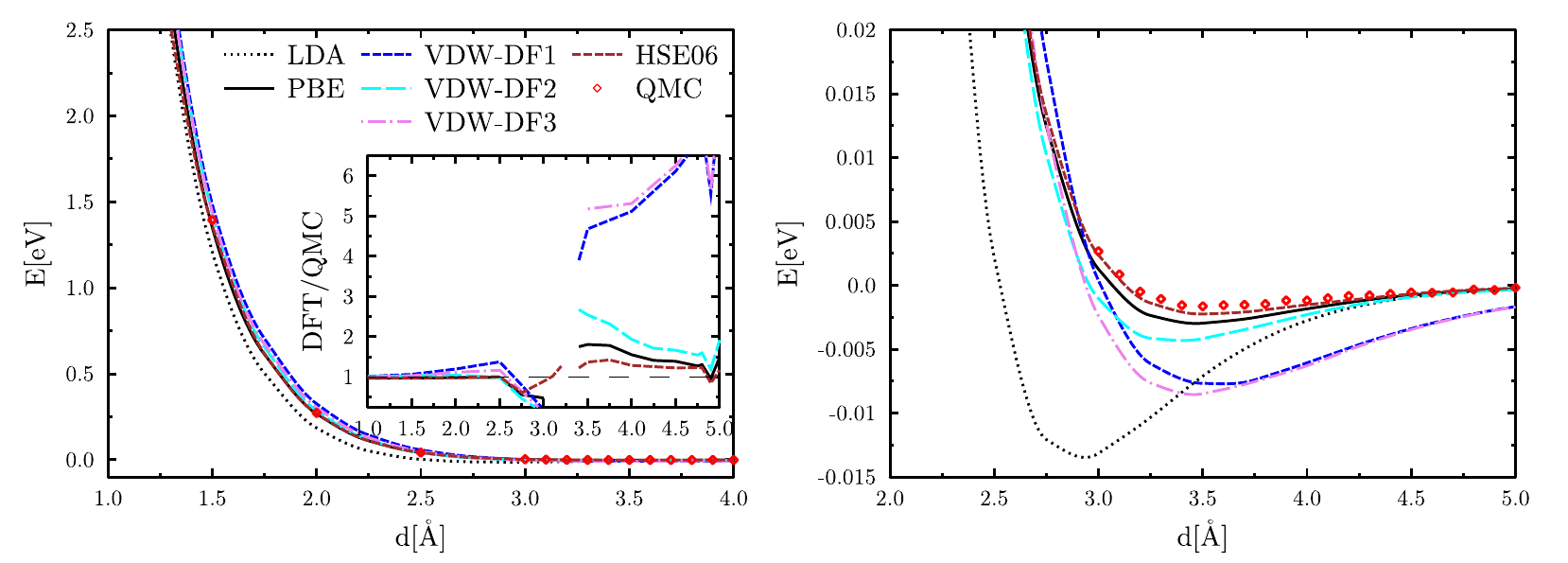}
     \put(85,19){\includegraphics[scale=0.5]{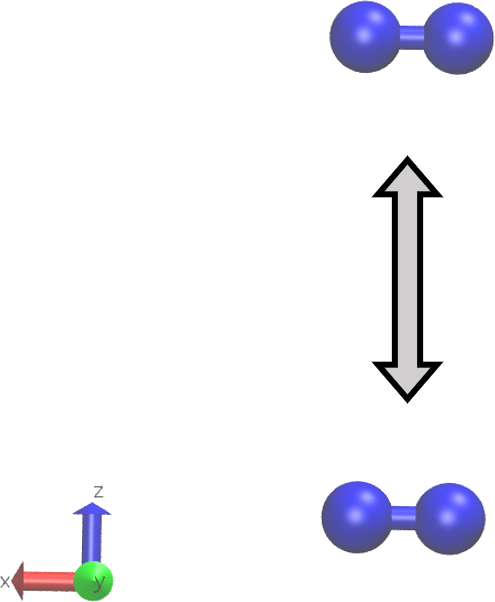}}
     \put(93.2,19){a}
     \put(92,25){d}
    \end{overpic}
    \caption{Model interaction geometry 2. The relative position of the four atoms ($L\ldots$ edge length of supercell, $d\ldots$ distance between molecules, $a\ldots$ molecule bond length) is shown in the top right corner: $(L/2,L/2,L/2-d/2)\,,(L/2+a,L/2,L/2-d/2)\,,(L/2,L/2,L/2+d/2)\,,(L/2+a,L/2,L/2+d/2)$.
    The left panel contains the overview and as the inset the relative deviation of the DFT results (featuring different xc functionals) from the QMC result (no curves are given in the inset for distances where the QMC energy is zero). The right panel shows the area of the vdW minimum of the interaction energy of the system of two molecules as function of the distance of their centre of mass. The energy at maximum intermolecular distance was taken as the zero point of the energy. QMC results contain standard errors no greater than 0.07 meV (smaller than the symbol size).}
    \label{fig:h2h2:2}
\end{figure*}
\begin{figure*}[t]
    \centering
    \begin{overpic}[width=\textwidth]
     {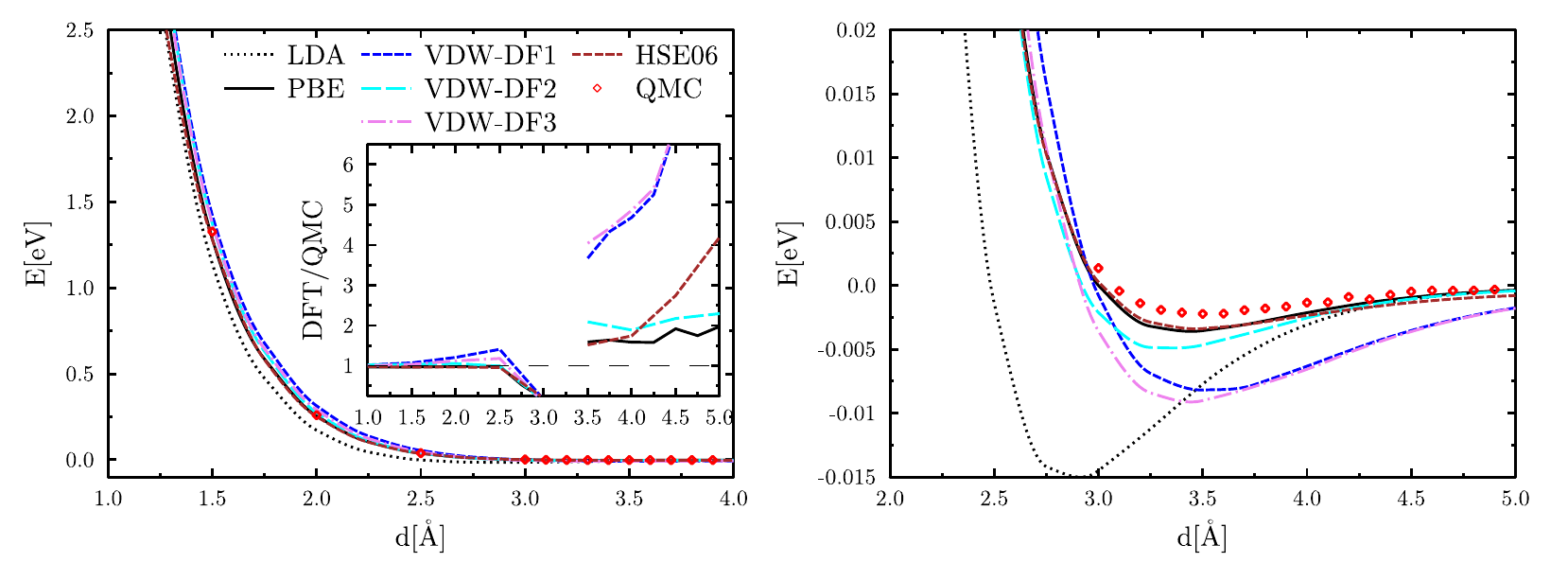}
     \put(82,17){\includegraphics[scale=0.6]{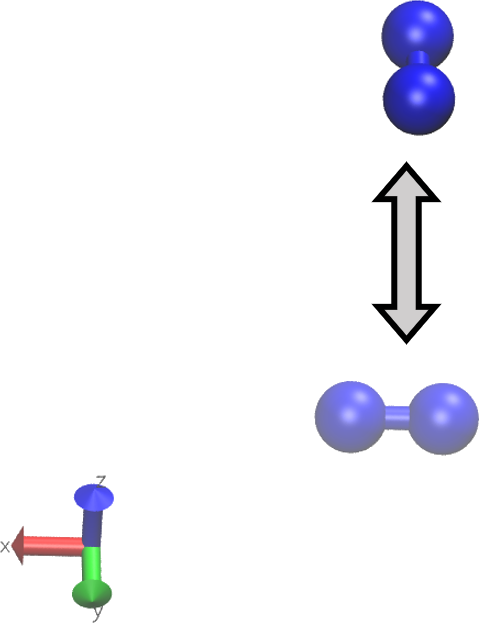}}
     \put(91.5,20){a}
     \put(90.5,26){d}
    \end{overpic}
    \caption{Model interaction geometry 3. The relative position of the four atoms ($L\ldots$ edge length of supercell, $d\ldots$ distance between molecules, $a\ldots$ molecule bond length) is shown in the top right corner: $(L/2-a/2,L/2,L/2-d/2)\,,(L/2+a/2,L/2,L/2-d/2)\,,(L/2,L/2-a/2,L/2+d/2)\,,(L/2,L/2+a/2,L/2+d/2)$.
    The left panel contains the overview and as the inset the relative deviation of the DFT results (featuring different xc functionals) from the QMC result (no curves are given in the inset for distances where the QMC energy is zero). The right panel shows the area of the vdW minimum of the interaction energy of the system of two molecules as function of the distance of their centre of mass. The energy at maximum intermolecular distance was taken as the zero point of the energy. QMC results contain standard errors no greater than 0.12 meV (smaller than the symbol size).}
    \label{fig:h2h2:3}
\end{figure*}
\begin{figure*}[t]
    \centering
    \begin{overpic}[width=\textwidth]
     {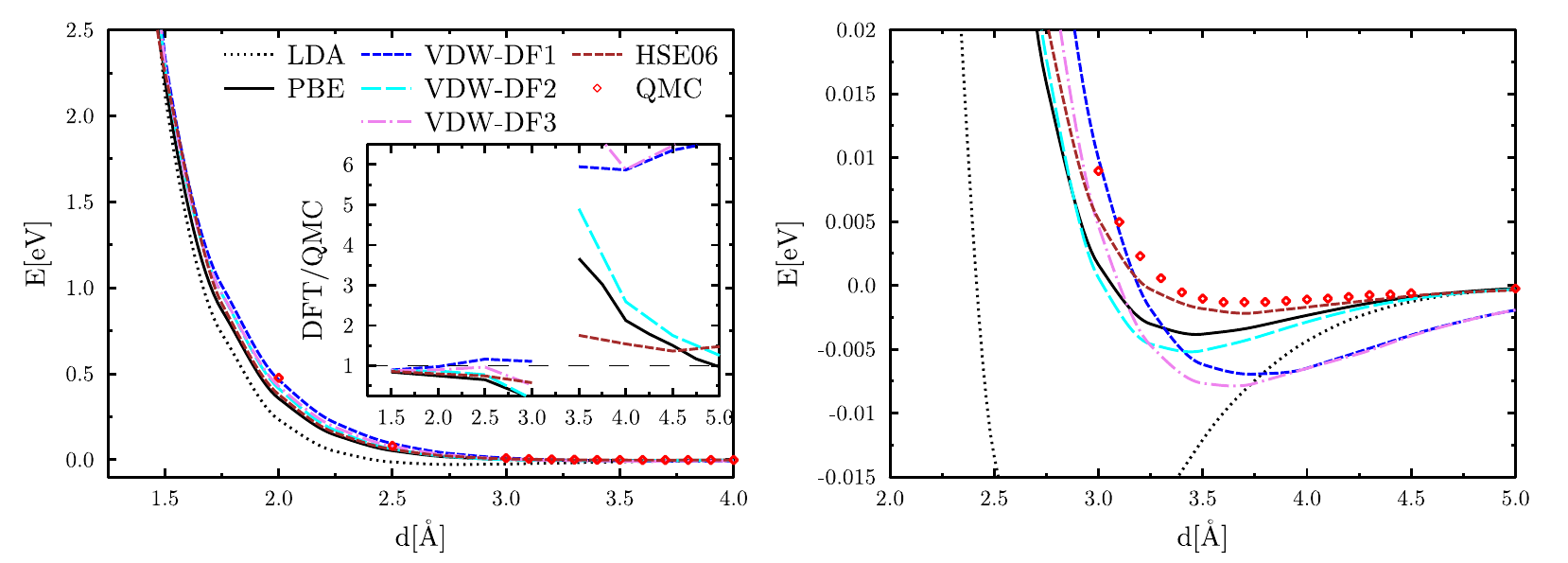}
     \put(82,18){\includegraphics[scale=0.6]{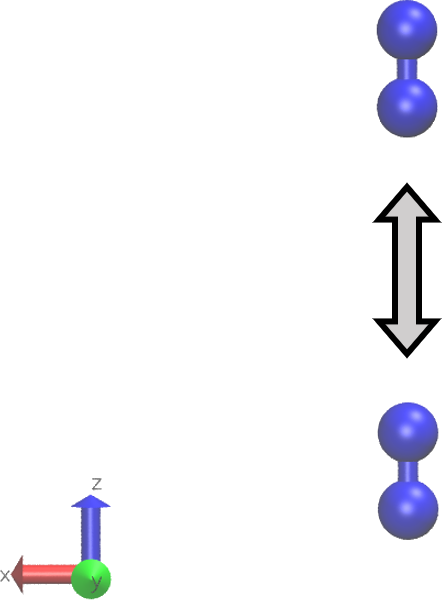}}
     \put(90.5,21){a}
     \put(90.5,26){d}
    \end{overpic}
    \caption{Model interaction geometry 4. The relative position of the four atoms ($L\ldots$ edge length of supercell, $d\ldots$ distance between molecules, $a\ldots$ molecule bond length) is shown in the top right corner: $(L/2,L/2,L/2-d/2)\,,(L/2,L/2,L/2-d/2+a/2)\,,(L/2,L/2,L/2+d/2)\,,(L/2,L/2,L/2+d/2+a/2)$.
    The left panel contains the overview and as the inset the relative deviation of the DFT results (featuring different xc functionals) from the QMC result (no curves are given in the inset for distances where the QMC energy is zero). The right panel shows the area of the vdW minimum of the interaction energy of the system of two molecules as function of the distance of their centre of mass. The energy at maximum intermolecular distance was taken as the zero point of the energy. QMC results contain standard errors no greater than 0.08 meV (smaller than the symbol size).}
    \label{fig:h2h2:4}
\end{figure*}

As the two-atom hydrogen molecule is of a noticeable size in comparison with typical distances of the vdW-minimum, we need to consider several different orientations of the two hydrogen molecules with respect to each other in order to get a general impression of the performance of the vdW functionals. This is explored in Figs.~\ref{fig:h2h2:2}, ~\ref{fig:h2h2:3}, and \ref{fig:h2h2:4}. The relative positions and orientations are always given in the captions of the figures.
The location of the minimum depends on the chosen geometry of the setup, the change is consistent with all xc functionals. As in Fig.~\ref{fig:h2h2:1}, we observe that VDW-DF2 is close to the PBE result, albeit with a slightly deeper minimum. VDW-DF1 and VDW-DF3 always show a minimum further out with a deeper well. 
We can summarize that in such an idealized situation, HSE06 gives best agreement with QMC when investigating vdW effects. PBE is a close second in the description of vdW effects and is on par with HSE06 when it comes to molecular dissociation energies. Comparing the computational cost of HSE06 and PBE, PBE seems clearly the best choice for the xc functional in warm dense hydrogen when vdW effects are of interest. It is worth pointing out that the vdW-minimum is of order $5$~meV and therefore a tiny contribution to the total energy with a typical scale in the single to multiple eV range.
\subsection{EOS using van-der-Waals functionals}\label{vdweos}
\begin{figure*}[t]
    \centering
    \includegraphics[width=\textwidth,clip=true]{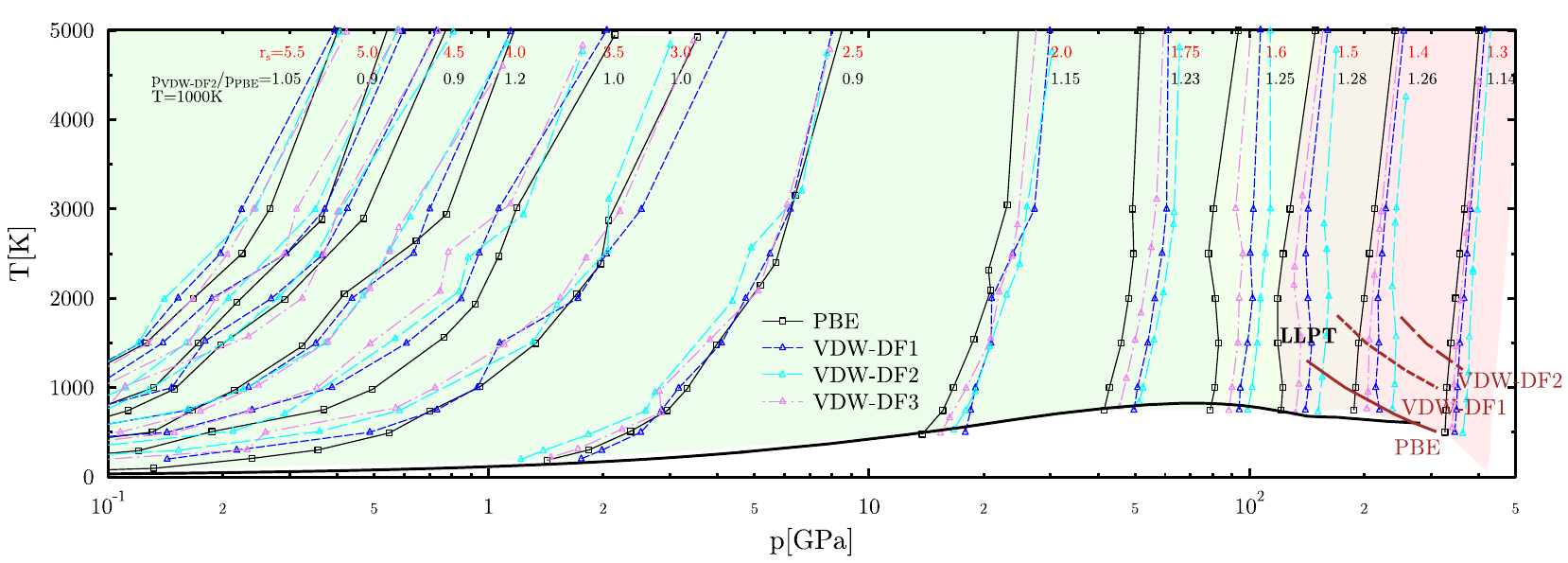}
    \caption{The investigated phase space of the hydrogen EOS using different xc functionals. At the top, the densities belonging to each isochor are indicated in magenta. In the line below (cyan color) the relative difference of the pressure at $T=1000$~K as obtained with PBE and VDW-DF2, resp., are shown. The hydrogen melt line (solid black curve) and several calculated phase transition lines of the LLPT (brown lines) are indicated~\cite{Diatschenko_1981,Lorenzen2010,Knudson_Science_2015,Lu_2019}.}
    \label{fig:eos}
\end{figure*}
For realistic scenarios of typical warm dense or high pressure hydrogen, resp., we study an array of densities and temperatures that span the ideal molecular regime, interacting (stable) molecular regime, and the parameter space where pressure dissociation occurs for temperatures where molecules should be stable. An overview of this parameter space is given in Fig.~\ref{fig:eos}. In addition to the isotherms, the relative deviation of the VDW-DF2 functional based EOS from PBE at $T=1000$~K is given. Deviations of the EOS' of different xc functionals at low densities (high $r_s$) are not an expression of different physics being described but are mainly due to inherent uncertainties in the DFT and MD parts at low and very low pressures and represent the obtainable precision in DFT-MD simulations within reasonable compute time. At high densities (low $r_s$) however, a significant trend of the vdW functionals to predict increased pressure values is observed in the region where molecules strongly interact and pressure dissociation occurs. This is consistent with our earlier argument about the vdW functionals producing substantially more stable molecules with higher dissociation energies (higher than PBE and higher than experiment), thus requiring more pressure work to break molecules.
\subsection{Structure using van-der-Waals functionals}\label{vdwsk}
\begin{figure*}[t]
    \centering
    \includegraphics[width=\textwidth,clip=true]{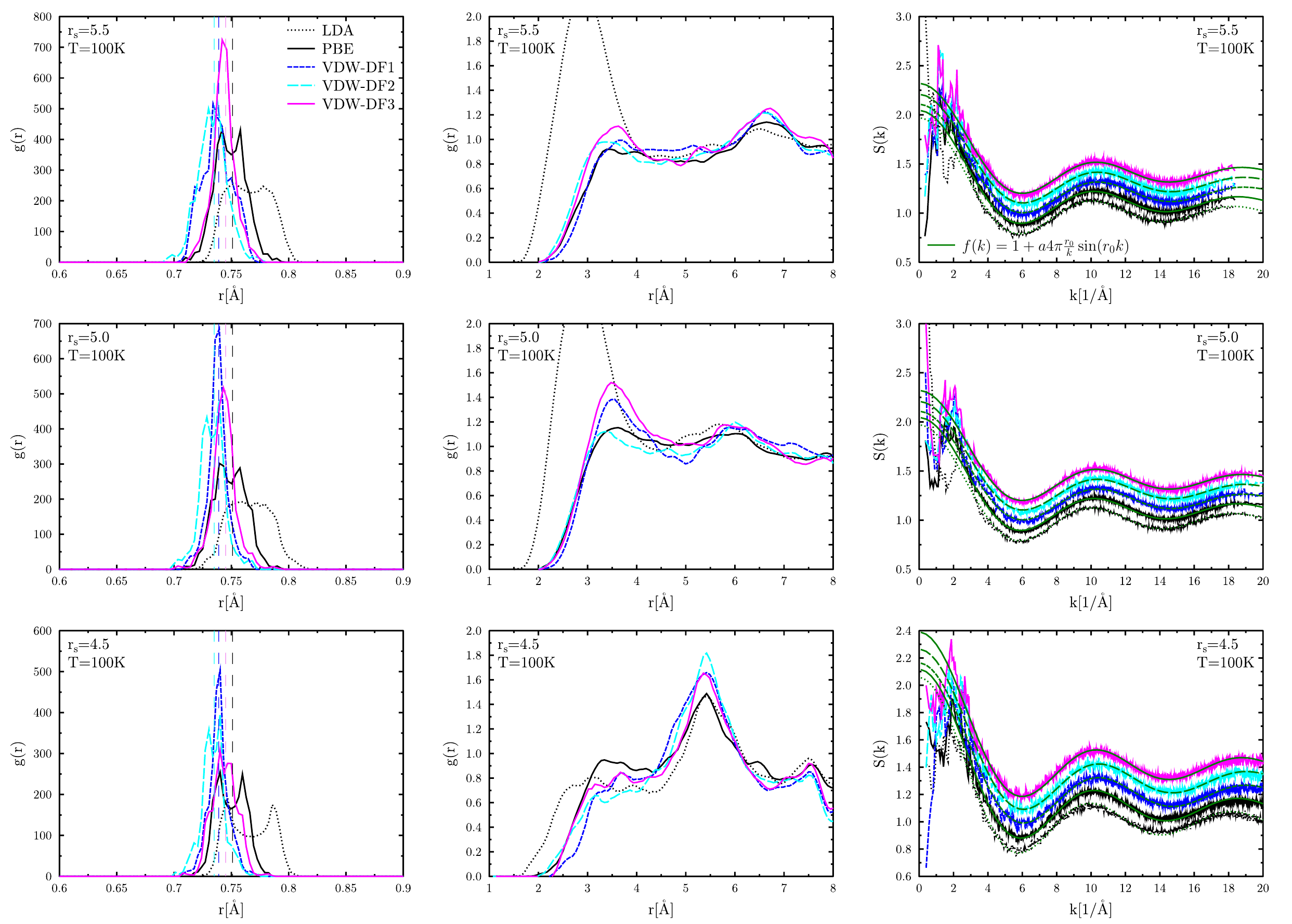}
    \caption{Hydrogen structure at $T=100$~K. The left column shows the molecular peak within the pair correlation function. The middle column shows the vdW region of the pair correlation function. The right column shows the static ion structure factor. The different structure factors are each shifted vertically by $0.1$ starting with LDA. The green lines are fits to the structure factor using the model function as indicated.}
    \label{fig:gr:100}
\end{figure*}
We start our analysis of the molecular and vdW structure, and its dependence on the xc functional in a many-particle system at low temperatures and low densities where the interacting molecules should most closely resemble the ideal molecules studied in the preceding sections. Figure~\ref{fig:gr:100} shows the ionic pair correlation function and static structure factor for hydrogen at $T=100$~K. We observe that the molecular peak in the pair correlation (left column of Fig.~\ref{fig:gr:100}) has a height and fine structure that depend on the xc functional. In particular, the LDA and PBE pair correlation functions feature a two-peak molecular maximum whereas VDW-DF1 and VDW-DF3 show a single peak only. This is contrary to the results we have shown earlier, where the potential energy curves that get established between the two protons that form the molecule do not differ at all at energy scales corresponding to temperatures of a few hundred Kelvin.

A temperature of $T=100$~K corresponds to an energy of several meV, thus is on the order of vdW interactions. The middle column of Fig.~\ref{fig:gr:100} focuses on the part of the pair correlation function around a few \AA, where the second peak in the pair correlation function, that describes the attraction of two molecules due to the vdW potential minimum, is located. As we have seen in the idealized scenarios of two-molecule-interactions, the height of these second peaks depend on the xc functional. Since we were able to demonstrate that PBE and VDW-DF2 match QMC data best, and we observe similar comparative behavior here, we should conclude that the vdW-peaks predicted by VDW-DF1 and VDW-DF3 are exaggerated. 

Comparing the weight of the molecular peak with the tiny deviations from unity that are caused by the vdW interactions visible in the pair correlation functions (middle column in Fig.~\ref{fig:gr:100}), it is clear that vdW interactions do not contribute in any measurable way to the physics of the system and any differences in predictions for the EOS or other quantities are rather a consequence of the different (not better) description of the molecular properties by the vdW xc functionals. 

In the right column of Fig.~\ref{fig:gr:100}, we show the static structure factors at these conditions. As the molecular peak is the dominant part in the pair correlation function, the static structure factor basically shows a damped oscillation whose wavelength is given by the bond length of the molecule. A fit of the following form is prudent if one considers the Fourier transform of a $\delta$-peak
\begin{align}
    S(k)=1+4\pi a\frac{r_0}{k}\sin{r_0k}\,,
\end{align}
where $a$ is a fit constant and $r_0$ is the bond length of the molecule. As can be seen, the different bond lengths due to different xc functionals are nicely represented in the static structure factors.

\begin{figure*}[t]
    \centering
    \includegraphics[width=\textwidth,clip=true]{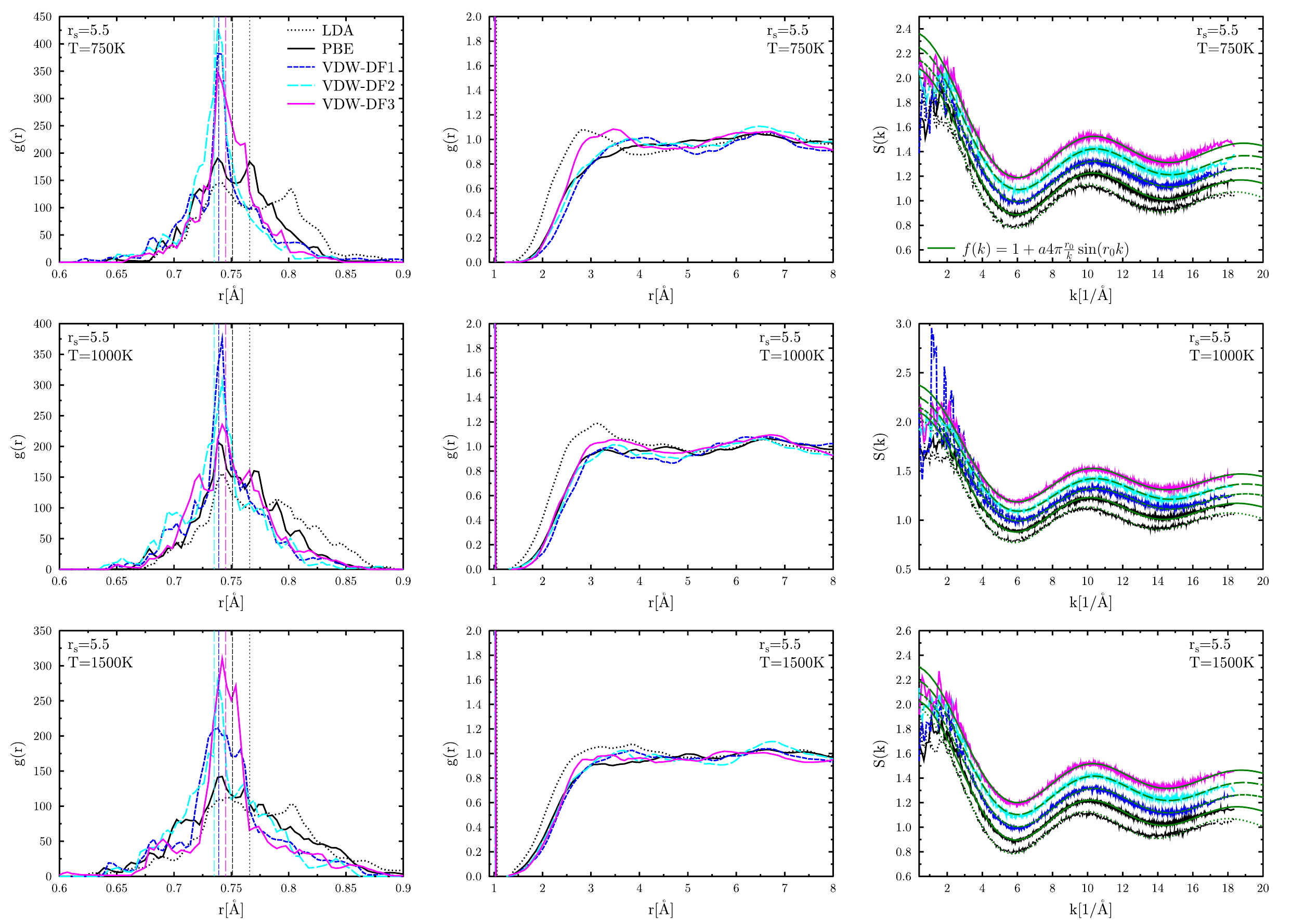}
    \caption{Hydrogen structure at $r_s=5.5$. The left column shows the molecular peak within the pair correlation function. The middle column shows the vdW region of the pair correlation function. The right column shows the static ion structure factor.  The different structure factors are each shifted vertically by $0.1$ starting with LDA. The green lines are fits to the structure factor using the model function as indicated.}
    \label{fig:gr:5.5}
\end{figure*}
Figure~\ref{fig:gr:5.5} shows our results for the lowest considered density of $r_s=5.5$. The three temperatures are a typical energetic regime of the LLPT. At these very much lower densities, molecules are dominating and are very stable. The left column of Fig.~\ref{fig:gr:5.5}, portraying the molecular peak, again shows fine structure and height differences between the different xc functionals. These are indications for different stabilities and energetic realities of the molecules. 

Contrary to the $T=100$~K case discussed earlier, there are no differences in the pair correlation functions at intermediate distances at relevant temperatures (neglecting LDA). vdW interactions thus do not play a role here. The different bond lengths of the molecules subject to different xc functionals are again nicely visible in the oscillations of the static structure factor shown in the right column of Fig.~\ref{fig:gr:5.5}.

\begin{figure*}[t]
    \centering
    \includegraphics[width=\textwidth,clip=true]{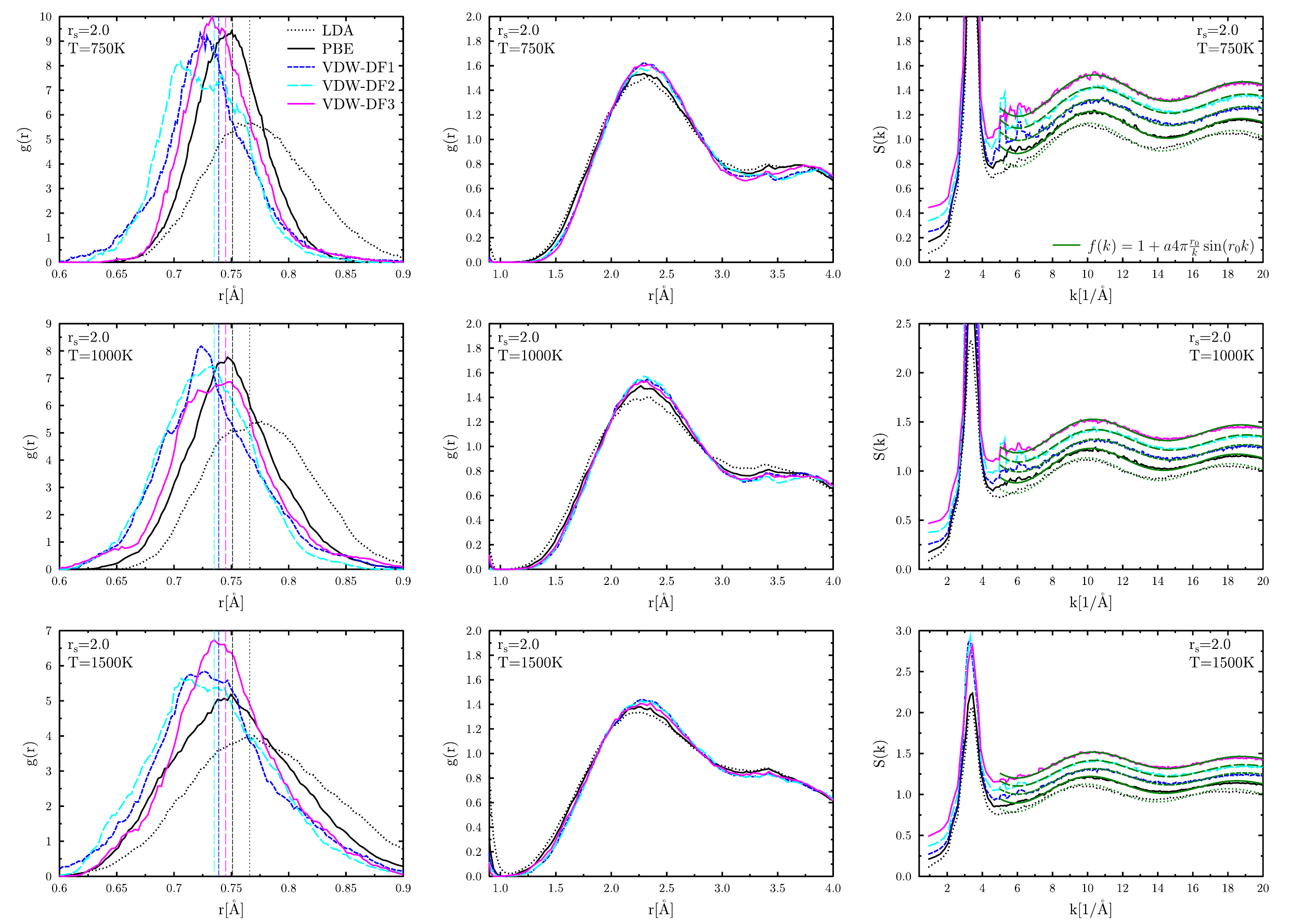}
    \caption{Hydrogen structure at $r_s=2.0$ for three different temperatures. Panel ordering as above.}
    \label{fig:gr:2.0}
\end{figure*}
We next consider a higher density of $r_s=2.0$, which is still well within the molecular phase, even though the interactions of the molecules influence the behavior. This situation is depicted in Fig.~\ref{fig:gr:2.0}. At once, the difference in position and shape of the molecular peak are obvious. All molecular peaks now resemble bell shaped curves without double peaks and fine structure. The second peak of the pair correlation function is well visible and indicates short range ordering. If there are differences at all in the second peak for different xc functionals, they are very small. The short range order is also visible in the static structure factor as there is a prominent first peak. The typical molecular oscillations only dominate the static structure for wavenumbers larger than $k=5$\AA$^{-1}$.

\begin{figure*}[t]
    \centering
    \includegraphics[width=\textwidth,clip=true]{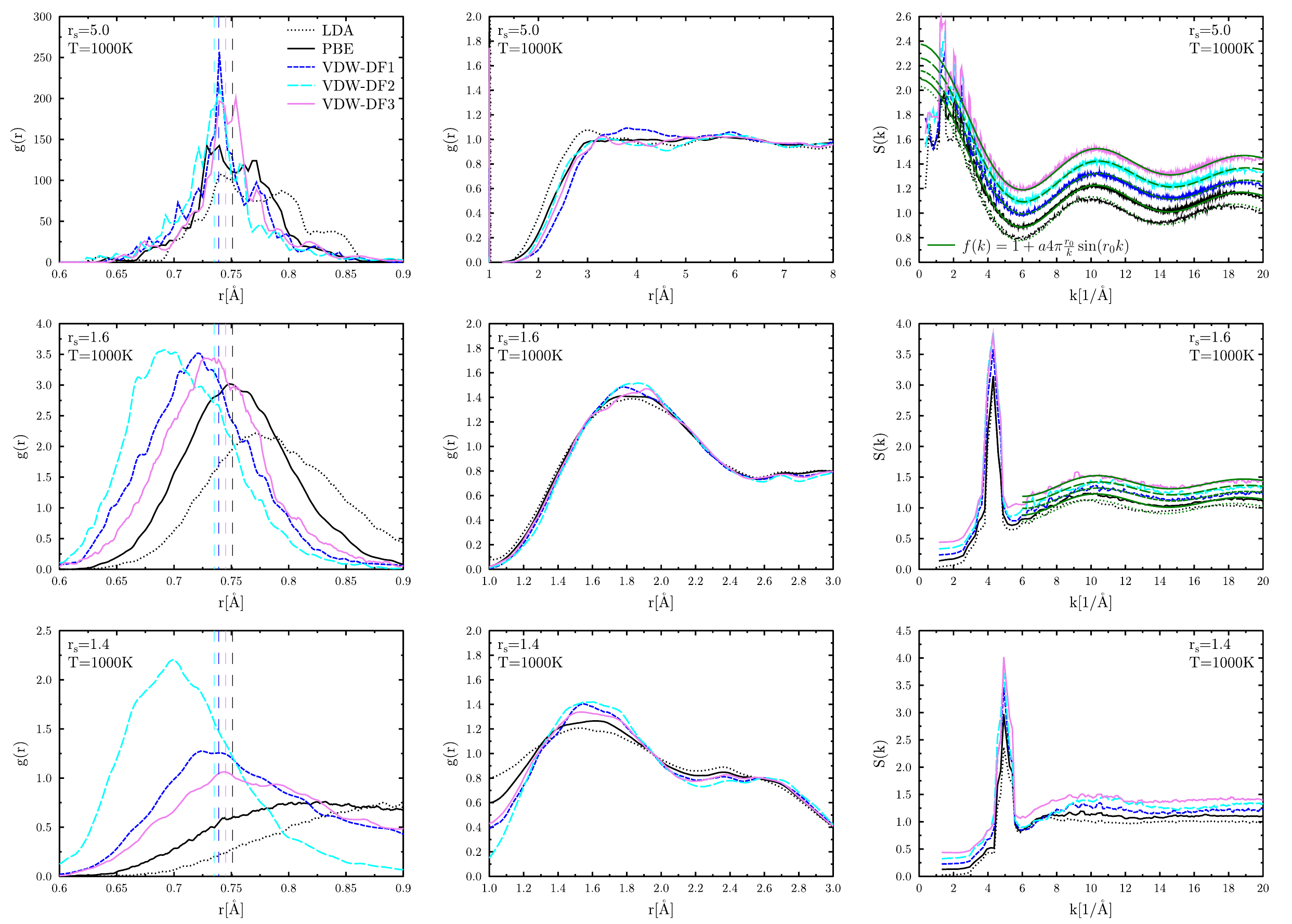}
    \caption{Static hydrogen structure at $T=1000$~K for three different densities in the molecular and metallic regime.}
    \label{fig:gr:1000}
\end{figure*}
We study the density effects on the molecules in hydrogen at the interesting temperature of $T=1000$~K in Fig.~\ref{fig:gr:1000}. Three different densities are considered. The first, $r_s=5.0$, is deep in the low density region with stable molecules. This means, there is a fine structure on the molecular peak (double maxima), there is no second peak as the temperature dominates any vdW interactions, and a purely oscillating static structure factor. A density of $r_s=1.6$ (2nd row) is at the border towards pressure dissociation and thus the molecular peak, even though it is still the highest peak in the pair correlation, is widened considerably and shrunk in height drastically. The shift to smaller values of the bond lengths from their ideal, vacuum levels is easily observed and also the dependence of this behaviour on the employed xc functional. The shape and location of the second peak, even though basically independent of the xc functional, is important and represents short range order when neighboring molecules feel their presence and interact. Finally, at $r_s=1.4$, we find substantially different pair correlation functions. The vdW xc functionals, as they predict a larger molecular dissociation energy, still feature molecules in the simulation at these conditions. PBE and LDA predict that all the molecules are very unstable and are being dissociated and their pair correlation functions and structure factors are those of a fluid at the verge of being metallic.

\begin{figure*}[t]
    \centering
    \includegraphics[width=\textwidth,clip=true]{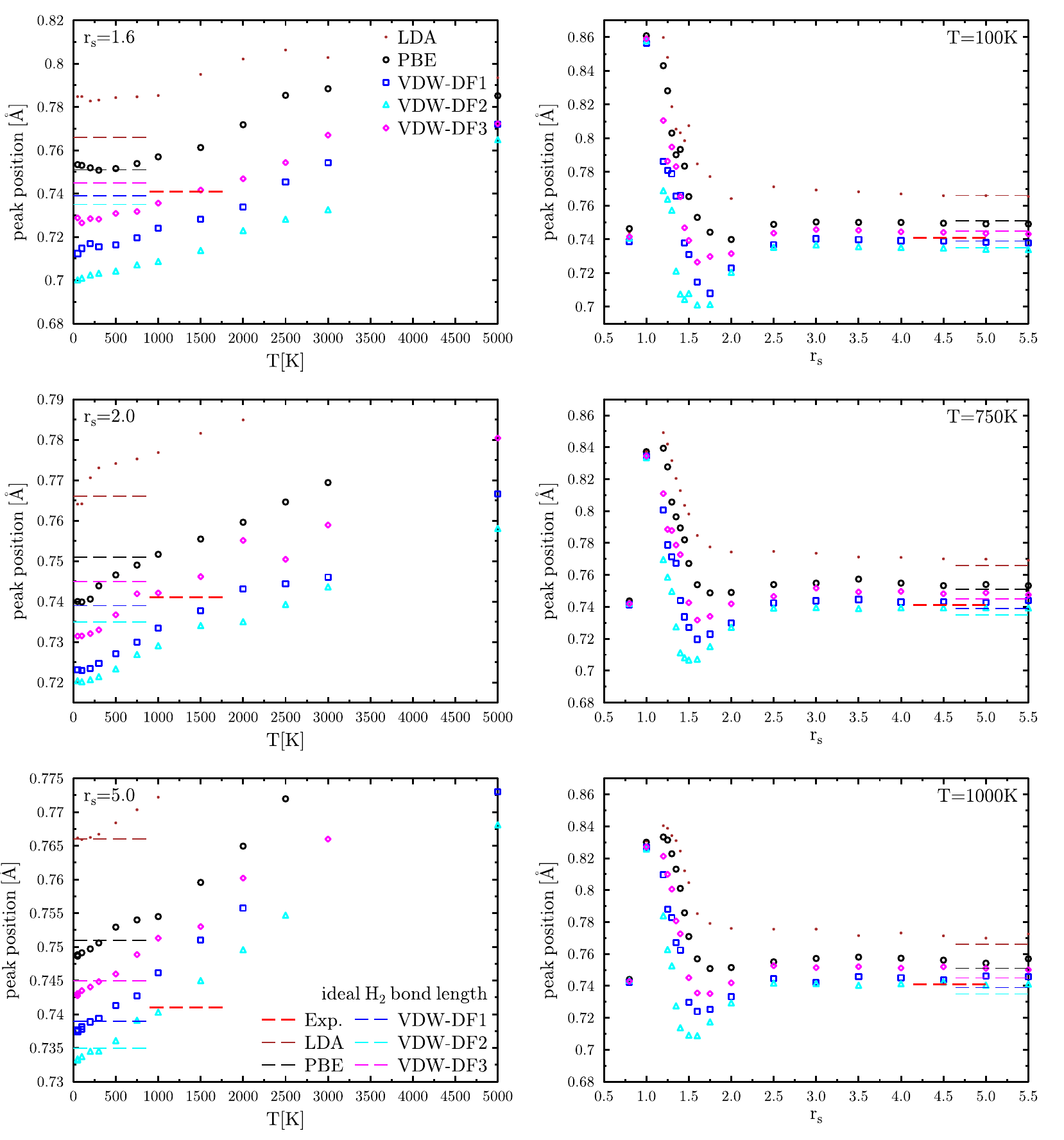}
    \caption{The center-of-mass peak positions of the molecular peak in hydrogen. The left column displays the trends with respect to temperature for three different densities. The right column shows the behaviour with respect to the density for three different temperatures. Thin dashed horizontal lines indicate the ideal $T=0$ bond lengths for the different xc functionals (color coded).}
    \label{fig:gr:pp}
\end{figure*}
In order to study the change of the molecular bond length with temperature and density, we first need to introduce a generalized measure for the bond length. As we have seen, e.g., in Figs.~\ref{fig:gr:100} \& \ref{fig:gr:5.5}, the molecular peak of the pair correlation function has a fine structure and may feature multiple peaks. Therefore, merely taking the location of the maximum value of the pair correlation function seems insufficient. We define the center-of-mass peak position $p_p$ and therefore the effective bond length of the hydrogen molecule as
\begin{equation}
    p_p=\frac{\sum\limits_{\substack{i=m \\ r_a < r_i < r_b}}^{n} r_i g(r_i)}
    {\sum\limits_{\substack{i=m \\ r_a < r_i < r_b}}^{n} g(r_i)}\,.
\end{equation}
The lower limit $0.4\le r_a \le 0.6$~\AA\; and the upper limit $0.9\le r_b \le 1.2$~\AA\; are varied to include the entire molecular peak (up to the minimum between the molecular peak and the continuum) and to achieve convergence.
The results of this analysis is presented in Fig.~\ref{fig:gr:pp}. In order to compare with the analysis of the ideal molecules in section~\ref{bl}, we have plotted thin dashed horizontal lines in the panels that indicate the bond length of the ideal molecules for the different xc functionals. As can be seen in the right column for large $r_s$ (small densities), the center-of-mass positions are an excellent tool to determine the bond length in the many-particle hydrogen systems and agree very well with the ideal bond lengths.

The density dependence of the bond lengths for all temperatures considered (right column of Fig.~\ref{fig:gr:pp}) is almost the same: Up to about $r_s=2.5$ there is no dependence on the density for $T=100$~K as an increase in density at these low densities only means that the molecules move slightly closer, but essentially remain independent. For $T=750$~K and $T=1000$~K, there seems to be a very slight maximum in the peak position around $r_s=3.2$. For densities higher than $r_s=2.5$, the bond lengths shorten, the molecules get compressed and become smaller. This effect is larger for the VDW xc functionals and almost absent for LDA. Once the minimum bond length (minimum in the center-of-mass peak position) is reached at around $r_s=1.6$ to $r_s=1.5$, the peak position becomes considerably larger. At this point very many molecules pressure dissociate and it becomes doubtful whether the first peak is actually still a molecular peak. The order of the curves remains the same throughout, i.e., PBE always predicts a larger bond length (peak position) than the vdW xc functionals.

Turning now to the temperature dependence of the bond length at constant densities, the left column of Fig.~\ref{fig:gr:pp} reveals that the bond length increases with temperature monotonically. This is due to the asymmetric (anharmonic) nature of the energy minimum between the two protons that determine the bonding, see Fig.~\ref{fig:h2:m2a}. Again, the order of the bond length predicted by the different xc functionals never changes. The tiny reduction of the bond lengths/peak positions with respect to the ideal bond lengths at small temperatures has the same origin as already explained in the right column. Overall, the temperature changes the effective bond length (center of mass peak position of the molecular peak of the pair correlation function) in the considered temperature range by about $5$\% for all xc functionals tested. This change caused by the temperature is thus of the same order as the change due to density before pressure dissociation sets in, albeit with different sign. The main difference is that the temperature influence materializes a lot more gradual for even higher temperatures whereas the increasing density puts a hard limit on the stability of the molecules due to the overlap of molecular wavefunctions.
\subsection{Dynamic ion structure \& ion modes}\label{vdwskw}
\begin{figure*}[t]
    \centering
    \includegraphics[width=\textwidth,clip=true]{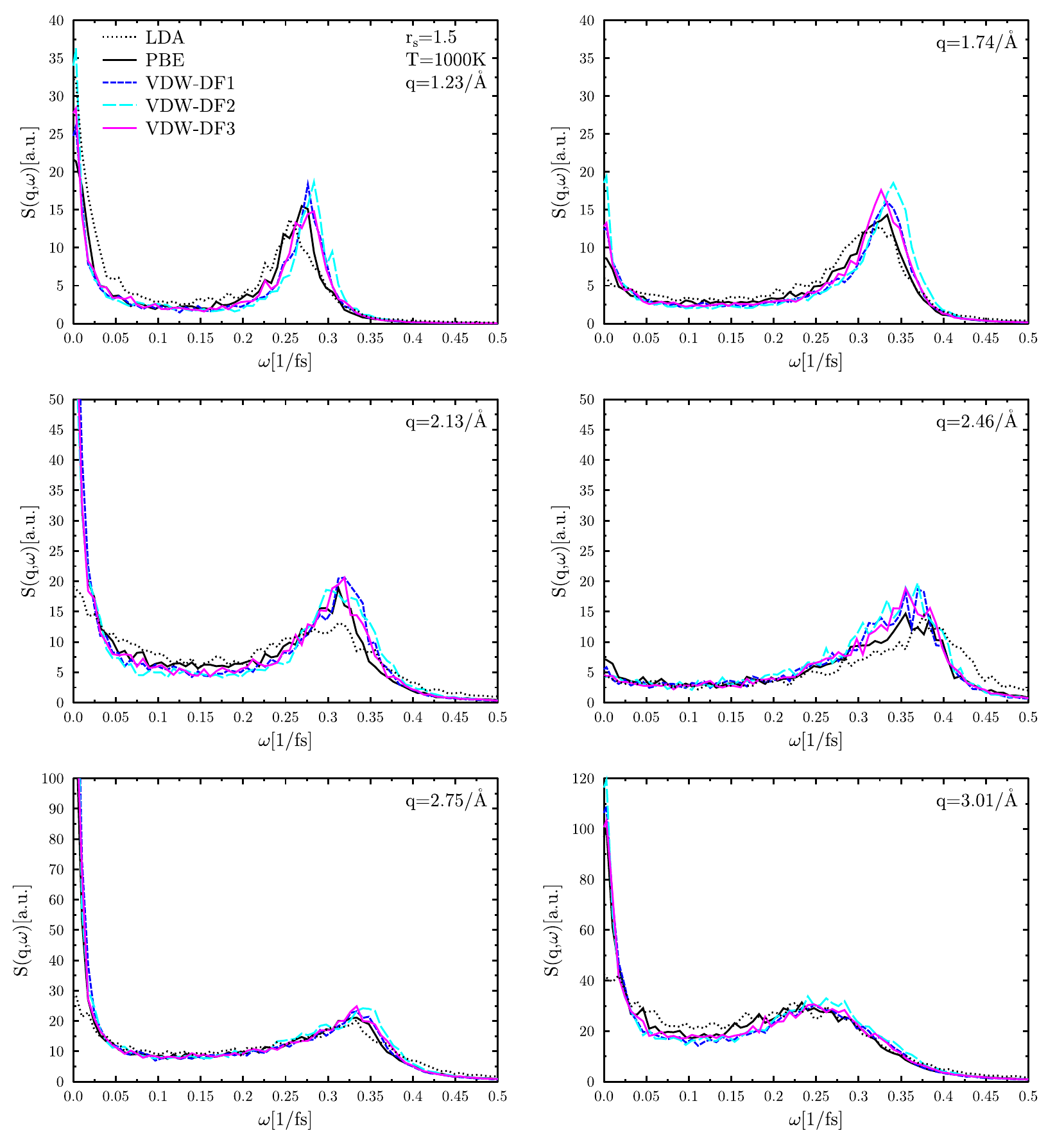}
    \caption{The ionic dynamic structure factor of hydrogen at $r_s=1.5$ and $T=1000$~K for six different wavenumbers.}
    \label{fig:skw_1.5}
\end{figure*}
Dynamic and collective effects in the ion structure can be investigated using the dynamic ion structure factor. We hope to observe signatures of different ion acoustic modes, different sound speeds, or different molecular oscillation frequencies depending on the xc functional.

We start the investigation of the dynamic ion structure at the density of $r_s=1.5$ and a temperature of $T=1000$~K, see Fig.~\ref{fig:skw_1.5}. This is a regime with severe differences in the pair correlation functions and static structure factors between calculations using different xc functionals. The molecules are highly unstable and, for PBE, on the verge of pressure dissociation. vdW molecules are still substantially more stable at these parameters. 

However, as can be seen seen in Fig.~\ref{fig:skw_1.5}, there are very little discernible differences in the dynamic ion structure factors for the first six wavenumbers that can be resolved in the simulation box. There is the ion acoustic peak starting at $0.275$~1/fs and moving toward larger frequencies before the ion acoustic peak location stays constant for further increased wavenumbers. At the largest plotted wavenumber, the location of the ion acoustic peak is getting smaller. The sharpness of the ion acoustic peak (the FWHM) is declining (increasing) all the time. The dispersion of the ion acoustic peak is therefore as expected even if in detail worthy of further investigation. The elastic part of the dynamic structure factor (at small frequencies) is increasing in weight with increasing wavenumber.

We do not observe any significant differences in the dynamic ion structure factor with reference to the xc functional. Even though we know from the analysis of the pair correlation function that there are significant differences in the amount of molecules and the ordering in the range up to $1$~\AA, these differences do not affect collective behavior in the hydrogen system. The reason for this is that the displayed wavenumbers show the collective regime (corresponding to mid- and long-range structural order) that is influenced by molecule-molecule interactions. This is precisely the regime where vdW interactions and their different treatment may be important. However, as, similarly to the mid-range pair correlation functions of, e.g., Figs.~\ref{fig:gr:2.0} \&~\ref{fig:gr:1000}, there is little to no difference in the dynamic structure factors, the molecule-molecule interactions are described almost identically using all different xc functionals and vdW effects are almost negligible.

\begin{figure*}[t]
    \centering
    \includegraphics[width=\textwidth,clip=true]{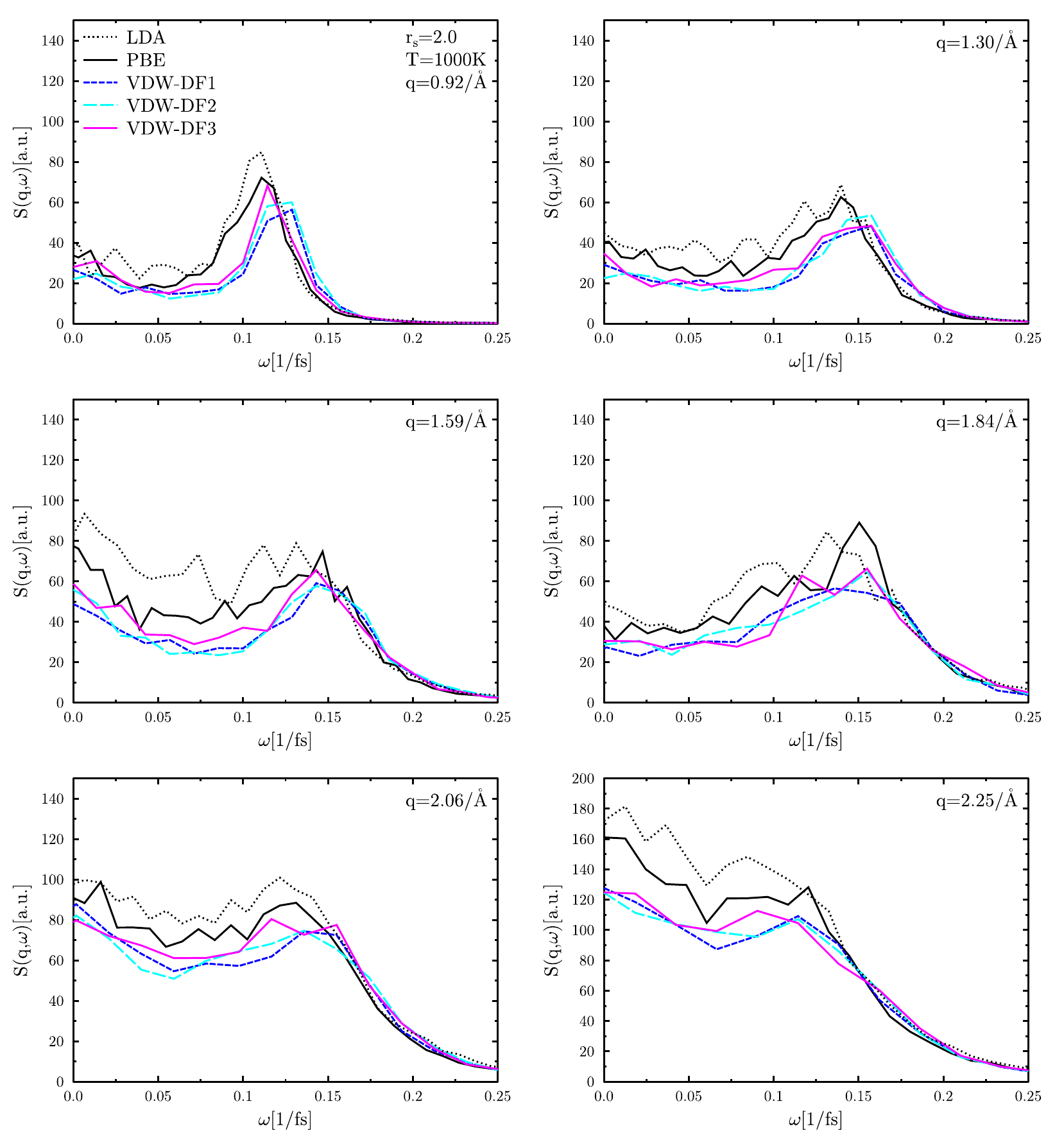}
    \caption{The ionic dynamic structure factor of hydrogen at $r_s=2.0$ and $T=1000$~K for six different wavenumbers.}
    \label{fig:skw_2.0}
\end{figure*}
The ion dynamic structure factor for a molecular hydrogen system featuring inter-molecular correlations at $r_s=2$ is given in Fig.~\ref{fig:skw_2.0}. The same physical situation was already explored and the static quantities shown in Fig.~\ref{fig:gr:2.0}. In the latter figure, we observed differences in the location of the molecular peak in the pair correlation function in accordance with the different bond lengths as predicted by different xc functionals. The second peak (vdW peak) was predicted to be the same using every xc functional. Here, in the dynamic ion structure factor, differences in the peak position of the acoustic mode can be seen only at the two smallest wavenumbers. Minor differences in the absolute value of the dynamic ion structure factor for smaller frequencies cannot be considered significant. Again, we point out that the shape of the dynamic structure factor at these wavenumbers is sensitive to mid- and long-range correlations which are described quite similarly using the different xc functionals.

\begin{figure*}[t]
    \centering
    \includegraphics[width=\textwidth,clip=true]{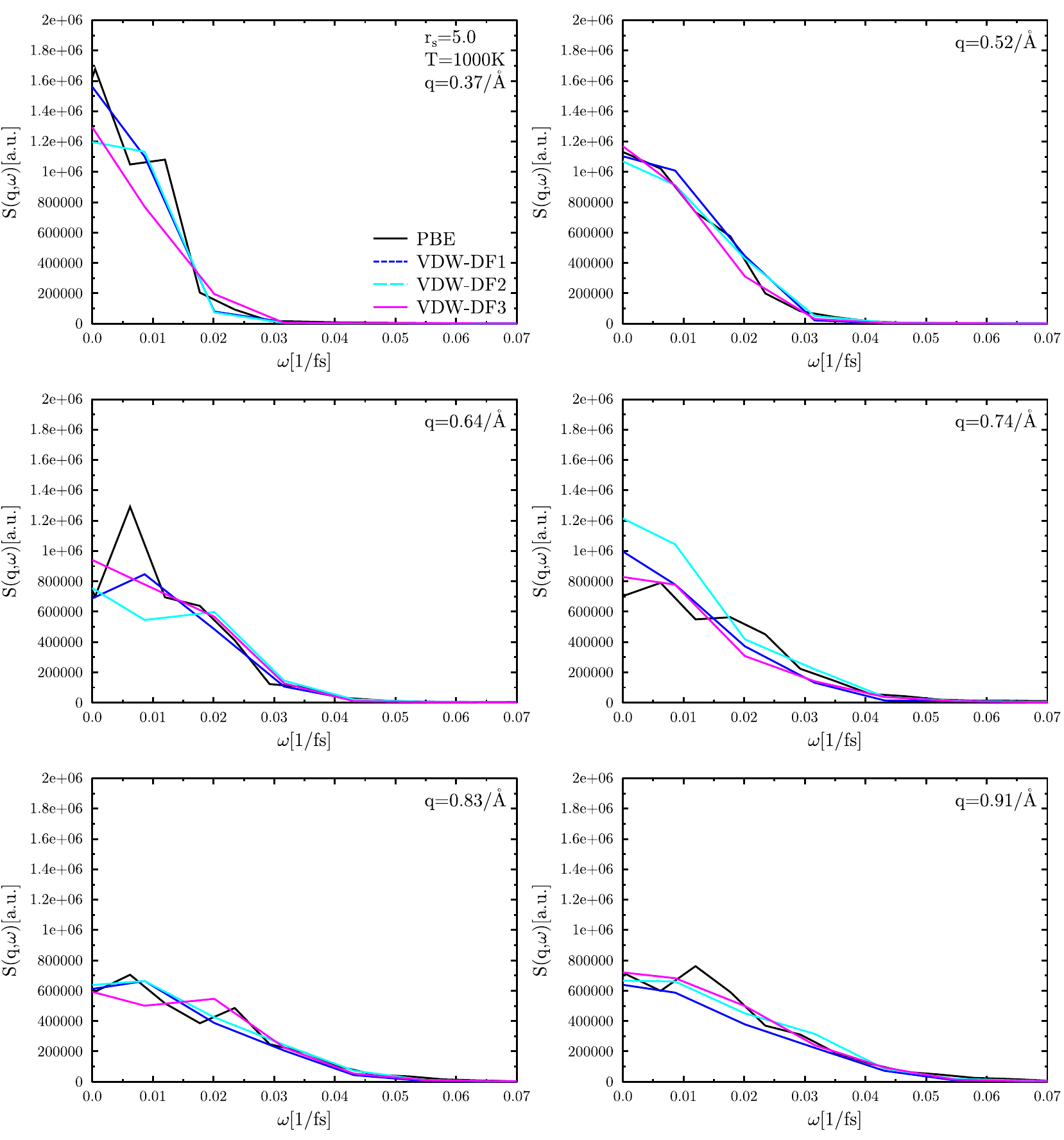}
    \caption{The ionic dynamic structure factor of hydrogen at $r_s=5.0$ and $T=1000$~K for six different wavenumbers.}
    \label{fig:skw_5.0}
\end{figure*}
The situation as depicted in Fig.~\ref{fig:skw_5.0} for a low density molecular system of almost no inter-molecular correlations is a very simple one. We do not observe collective modes, just a simple monotonic decay of the ion dynamic structure factor. More importantly, there are no differences due to different xc functionals. All molecules are sufficiently stable such that tiny differences in bond energies and bond lengths are washed out due to the temperature and the ion dynamics look the same, no matter the xc functional.
\subsection{Electronic structure}\label{vdwdos}
\begin{figure}[t]
    \centering
    \includegraphics[width=0.5\textwidth,clip=true]{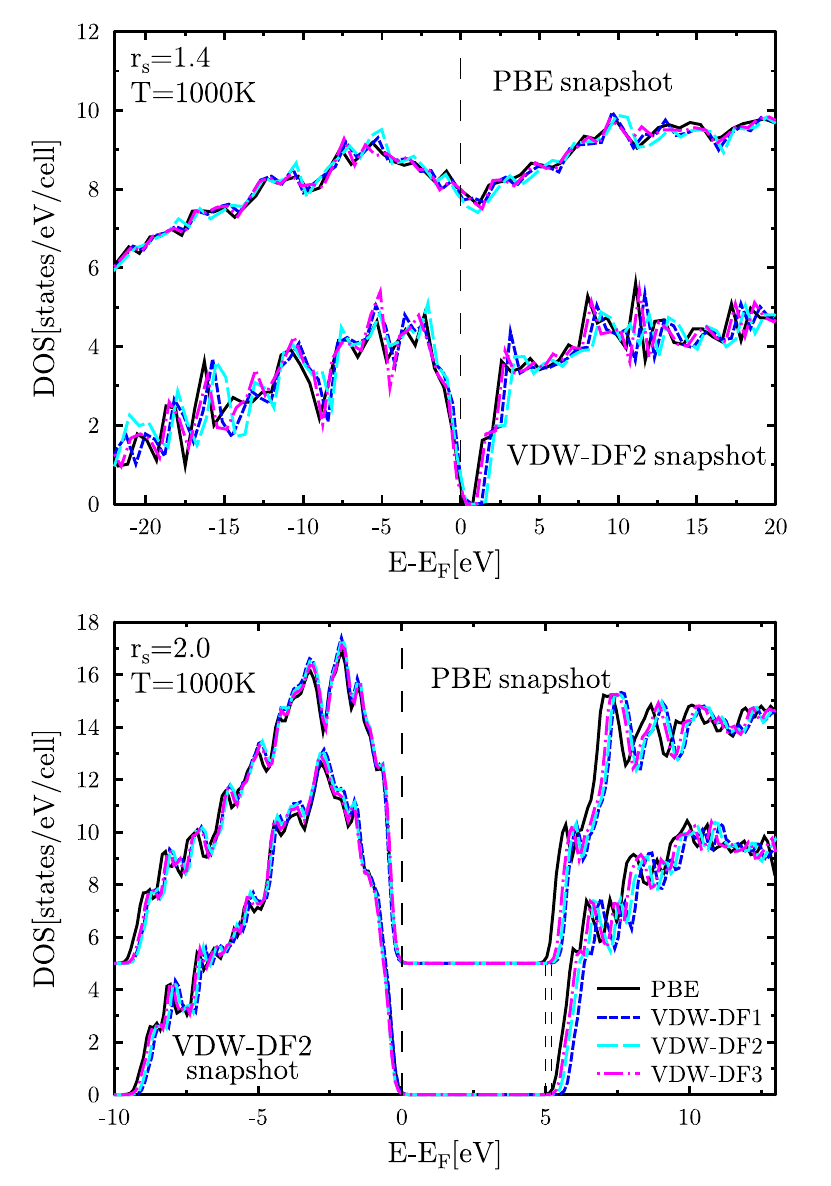}
    \caption{The electronic DOS of hydrogen for selected temperature and density conditions.The top panel shows a situation where PBE predicts a metallic state but the VDW functionals predict a molecular state. The bottom panel shows a molecular state.}
    \label{fig:edos}
\end{figure}
An important question in the physics of the molecular to metallic transition is whether it is an ionic structural change first (pressure dissociation of molecules) or an electronic change (closing of the band gap). In Fig.~\ref{fig:edos}, we illuminate the situation in the context of different xc functionals. In the bottom panel of Fig.~\ref{fig:edos}, we show the DOS for two different ionic snapshots at $r_s=2$ and $T=1000$~K. One configuration was obtained from a DFT-MD simulation using the PBE functional, the other one by a DFT-MD simulation using the VDW-DF2 functional. Both DOS show a HOMO-LUMO bandgap of about $5$~eV. This does not change when calculating the DOS for these snapshots using different xc functionals. The top four lines in the bottom panel of Fig.~\ref{fig:edos}, are the DOS when a DFT calculation using PBE, VDW-DF1, VDW-DF2, and VDW-DF3 functionals is performed on a PBE snapshot. The lower curves show the same situation for a VDW-DF2 snapshot. In all cases, the band gap remains stable of almost the same size.

The situation is a bit different for the higher density of $r_s=1.4$ at the same temperature, see top panel of Fig.~\ref{fig:edos}. The PBE simulations predict a dissociated metallic fluid, hence there is no bandgap. The VDW-DF2 simulations still show a small band gap of $0.5$~eV and there are still neutral molecules in the simulation box. These findings are again independent from the xc functional used to determine the DOS for the particular snapshot. The DOS depends almost entirely only on the ionic structure and only to a very small amount on the xc functionals.

\section{Summary}
We have performed a number of DFT and DFT-MD simulations using standard PBE, Meta-GGA, hybrid-GGA, as well as vdW functionals to elucidate their effect on a number of system properties with special attention to the question whether vdW effects cause the vastly different predictions for the LLPT in hydrogen at around $2$~Mbar. 

We investigated the molecular dissociation energy, the molecular bond length, and isolated molecule-molecule interactions for (a) molecule(s) in vacuum. We also looked at pair correlation functions and their fine structure, static and dynamic ion structure factors, bond length changes in correlated systems, the equation of state, and the electronic density of states. All of these quantities were compared for a standard PBE xc functional and for three different non-local vdW xc functionals.

In all of these quantities, there is no reason evident why vdW xc functionals should be trusted more or should give results superior to the PBE results. The molecular bond lengths predicted by vdW functionals are marginally better than the PBE ones. The dissociation energy of the hydrogen molecule is underestimated by PBE (compared to QMC or experimental values) but overestimated by a similar amount by the vdW functionals. The potential energy binding curves of the two protons that form the molecule are all but indistinguishable when adjusting for the different dissociation energies and all the DFT curves are in very good agreement with CISD results. Thus, we conclude that the higher transition pressure of the LLPT as reported using vdW xc functionals may only be due to the overestimated dissociation energy. 

Actual vdW effects in fluid hydrogen between two or more hydrogen molecules are better and more accurately described by PBE and HSE06 than by any vdW xc functional that we tested. In addition, the energy scale of polarization effects and neutral-neutral interactions ($\sim$meV) is vastly inferior to the energy scales of molecular dissociation or ionization ($\sim$eV) so that for relevant temperatures and pressures (for the LLPT) no difference in the structure of the hydrogen system at ranges of $1$~\AA~to $10$~\AA~could be detected. There were however differences in the fine structure of the molecular peaks in the pair correlation function and in the location of the same peak that can again be traced back, not to differences in the treatment of vdW-effects of the used xc functionals, but in their differences in describing the molecular bond in hydrogen.

Concluding that vdW effects matter less than initially believed for the EOS of liquid hydrogen under warm dense matter or high pressure conditions, we conjecture that the r$^2$SCAN functional is best (among the tested once) since, even though it gives terrible vdW energy curves, it matches the molecular dissociation energy and bond length best.

The main body of work concerning high pressure liquid hydrogen here and elsewhere has been dedicated to the EOS, the ion-ion and electron-ion structural properties. Future work therefore needs to be extended also to electron-electron correlation functions as a a way to improve our understanding of pressure dissociation. New quantum Monte Carlo methods or new DFT approaches should thus be applied~\cite{Dornheim2023,moldabekov2025abinitiodensityfunctional}.

\section*{Data Availability Statement}
The data that support the findings of this study are openly available in the Rossendorf Data Repository (RODARE) at \url{https://rodare.hzdr.de/}, in particular \url{https://doi.org/10.14278/rodare.4364}.

\section*{Acknowledgments}
Insightful discussions with Klaus Steiniger are gratefully acknowledged.

This work was partially supported by the Center for Advanced Systems Understanding (CASUS), financed by Germany’s Federal Ministry of Education and Research (BMBF) and the Saxon state government out of the State budget approved by the Saxon State Parliament.
This work has received funding from the European Research Council (ERC) under the European Union’s Horizon 2022 research and innovation programme
(Grant agreement No. 101076233, ``PREXTREME''). 
Views and opinions expressed are however those of the authors only and do not necessarily reflect those of the European Union or the European Research Council Executive Agency. Neither the European Union nor the granting authority can be held responsible for them. T.D.~gratefully acknowledges funding from the Deutsche Forschungsgemeinschaft (DFG) via project DO 2670/1-1.

The authors gratefully acknowledge the computing time granted by the Resource Allocation Board and provided on the supercomputer Emmy/Grete at NHR-Nord@G\"ottingen as part of the NHR infrastructure. The calculations for this research were conducted with computing resources under the project mvp00024. Computations were also performed on a Bull Cluster at the Center for Information Services and High-Performance Computing (ZIH) at Technische Universit\"at Dresden. We gratefully acknowledge compute time and data storage at the hemera cluster of the Department of Information Services and Computing of the HZDR.

This work was also supported in part by the U.S. Department of Energy, Office of Science, Office of Basic Energy Sciences Early Career Research Program (ECRP) under Award Number DE-SC0025900, in particular the research conducted by G.J.S., W.Z.V., H.R.P. and J.J.S. consisting of \emph{i}-FCIQMC, CCSD, and molecular DFT calculations.
This research also used computer resources from the University of Iowa and Michigan State University.

\bibliography{bibexport}
\end{document}